\documentclass[12pt]{iopart}
\usepackage{iopams}

\expandafter\let\csname equation*\endcsname\relax

\expandafter\let\csname endequation*\endcsname\relax

\usepackage{amsmath}

\usepackage{graphicx}
\usepackage{dcolumn}
\usepackage{bm}
\usepackage{hyperref}
\usepackage{xcolor}
\usepackage{layouts} 
\usepackage{tikz}
\usepackage{ifthen}
\usetikzlibrary{positioning, calc}
\usetikzlibrary{decorations.pathmorphing} 
\usepackage[export]{adjustbox}
\usepackage{xspace} 

\usepackage[backend=biber, style=numeric-comp, abbreviate=true, minbibnames=1, maxbibnames=3, alldates=iso, seconds=true, sorting=none,
isbn=false,
url=false,
eprint=false,
doi=false]{biblatex} 
\usepackage[inline,shortlabels]{enumitem}

\usepackage{booktabs} 
\usepackage{multirow}
\usepackage{bold-extra}
\newcommand{\tableheadline}[1]{\multicolumn{1}{c}{\bfseries #1}}

\usepackage{algorithm}
\usepackage{algpseudocode}

\usepackage{bbold}
\usepackage{soul}
\newcommand{\mathdefault}[1][]{} 
\newcommand{\avg}[1]{\langle {#1} \rangle}
\newcommand{\abs}[1]{\vert {#1} \vert}
\newcommand{\absbr}[1]{\left\vert {#1} \right\vert}
\newcommand{\bra}[1]{\langle {#1} \vert}
\newcommand{\ket}[1]{\vert {#1}\rangle}
\newcommand{\braket}[2]{\langle{#1} \vert {#2}\rangle}
\newcommand{\ketbra}[2]{\vert {#1} \rangle \langle{#2}\vert}

\DeclareMathOperator*{\argmax}{arg\,max}

\newcommand*{\ie}{i.e.\@\xspace}
\newcommand{\vect}[1]{\boldsymbol{#1}} 

\newcommand{\Op}{\Omega_\mathrm{p}}
\newcommand{\Opmax}{\Omega_\mathrm{p}^\mathrm{max}}
\newcommand{\Os}{\Omega_\mathrm{s}}
\newcommand{\Osmax}{\Omega_\mathrm{s}^\mathrm{max}}
\newcommand{\Ops}{\Omega_\mathrm{p/s}}
\newcommand{\Opsmax}{\Omega_\mathrm{p/s}^\mathrm{max}}
\newcommand{\ti}{t_\mathrm{i}}
\newcommand{\tf}{t_\mathrm{f}}
\newcommand{\delp}{\delta_\mathrm{p}}
\newcommand{\I}{\mathrm{i}}
\newcommand{\de}{\mathrm{d}}
\newcommand{\Hc}{H_\mathrm{c}}
\newcommand{\Hsys}{H_\mathrm{sys}}
\newcommand{\Hnoise}{H_\mathrm{noise}}
\newcommand{\Hnoisetilde}{\tilde{H}_\mathrm{noise}}
\newcommand{\xii}{\xi^{(r)}}
\newcommand{\psii}{\psi^{(r)}}
\newcommand{\xieq}{\xi_{\mathrm{p}=\mathrm{s}}}
\newcommand{\xigr}{\xi_{\mathrm{p}>\mathrm{s}}}
\newcommand{\xilo}{\xi_{\mathrm{p}<\mathrm{s}}}

\begin{document}

\title{Noise classification in three-level quantum networks by Machine Learning}

\author{Shreyasi Mukherjee$^1$, Dario Penna$^2$, Fabio Cirinnà$^2$, Mauro Paternostro$^{3,4}$, Elisabetta Paladino$^{1,5,6}$, Giuseppe Falci$^{1, 5, 6}$, and Luigi Giannelli$^{1,5,*}$}

\address{$^1$ Dipartimento di Fisica e Astronomia ``Ettore Majorana'', Università di Catania, Via S. Sofia 64, 95123 Catania, Italy}
\address{$^2$ Leonardo S.p.A., Cyber \& Security Solutions, 95121, Catania, Italy}
\address{$^3$ Universit\`a degli Studi di Palermo, Dipartimento di Fisica e Chimica - Emilio Segr\`e, via Archirafi 36, I-90123 Palermo, Italy}
\address{$^4$ Centre for Theoretical Atomic, Molecular, and Optical Physics, School of Mathematics and Physics, Queens University, Belfast BT7 1NN, United Kingdom}
\address{$^5$ INFN, Sezione di Catania, 95123, Catania, Italy}
\address{$^6$ CNR-IMM, UoS Università, 95123, Catania, Italy}

\ead{${}^*$luigi.giannelli@dfa.unict.it}

\vspace{10pt}

\begin{abstract}
    We investigate a machine learning based classification of noise acting on a small quantum network with the aim of detecting spatial or multilevel correlations, and the interplay with Markovianity. We control a three-level system by inducing coherent population transfer exploiting different pulse amplitude combinations as inputs to train a feedforward neural network. We show that supervised learning can classify different types of classical dephasing noise affecting the system. Three non-Markovian (quasi-static correlated, anti-correlated and uncorrelated) and Markovian noises are classified with more than $99\%$ accuracy. On the contrary, correlations of Markovian noise cannot be discriminated with our method. Our approach is robust to statistical measurement errors and retains its effectiveness for physical measurements where only a limited number of samples is available making it very experimental-friendly. Our result paves the way for classifying spatial correlations of noise in quantum architectures.
\end{abstract}

%
%

%
%
%


\section{\label{sec:Intro}Introduction}
Nowadays, quantum systems can be controlled with impressive accuracy~\cite{KochEQT2022quantum} to perform new tasks by exploiting their coherence properties, such as superposition and entanglement~\cite{AcinNJP2018quantum}. However, quantum behavior fades away because of environmental noise that leads to a loss of accuracy of quantum operations and decoherence~\cite{ZurekRMP2003decoherence}. 
The development of strategies to counteract these effects is therefore of crucial importance for progress in quantum technology. In this context, machine learning (ML) is proving to be an innovative and powerful diagnostic tool~\cite{MarquardtSPLN2021machine,KrennPRA2023artificial,GebhartNRP2023learning}. ML-based approaches have been applied to various quantum control protocols~\cite{BrownNJP2021reinforcement,GiannelliPLA2022tutorial,NiunQI2019universal,SgroiPRL2021reinforcement}, to quantum state tomography~\cite{BanchiNJP2018modelling,TorlaiNP2018neuralnetwork,PalmierinQI2020experimental}, to characterize quantum systems~\cite{CouturierE2023characterization,GenoisPQ2021quantumtailored,YoussrynQI2020characterization}, to simulate open quantum systems~\cite{BandyopadhyayCP2018applications,LuoPRL2022autoregressive,BanchiNJP2018modelling,YoussrynQI2020characterization}, to study non-Markovian dynamics~\cite{LuchnikovPRL2020machine,FanchiniPRA2021estimating} and, in a similar spirit similar to this work, to characterize system-environment interactions~\cite{PapicPRA2022neuralnetworkbased,WisePQ2021using} and perform noise spectroscopy~\cite{BarrMLST2024spectral,MartinaAQ2021learning,MartinaMLST2023deep}.

The mitigation of decoherence effects is necessary to achieve quantum advantage. Over the last two decades, several strategies have been developed for both for quantum computation and quantum control in the broader sense. Strategies for error avoidance (or passive stabilization) consist of storing and processing information in suitably designed subspaces of the Hilbert space which are protected from the interaction with the environment~\cite{PaladinoRMP2014oneoverf,VionS2002manipulating}. Examples of active stabilization strategies include Quantum Error Correction which relies on encoding nonlocal information~\cite{Roffe2019,Campbell2024,
}, and Dynamical decoupling~\cite{ViolaPRL1999dynamical,falci_dynamical_2004,suter_colloquium_2016,DArrigo2024openloop}, which consists in the repeated application of pulsed or switched control, which has been used in solid-state coherent nanoscience to counteract $1/f$ noise~\cite{PaladinoRMP2014oneoverf,falci_1f_2024,balandin_electronic_2024,pellegrino_1f_2020}.
Specific strategies rely on the availability of information about the environment, the characterization of which is therefore a very important step in the development of optimized systems and protocols. For single qubits, this programme has been successfully carried out in various platforms, such as trapped ions~\cite{Biercuk2009}, photonic qubits~\cite{Damodarakurup2009,orieux_experimental_2015}, nuclear magnetic resonance~\cite{Jenista2009,
}, and nitrogen vacancy centres in diamond~\cite{Naydenov2011}. Superconducting systems are an outstanding example in this context~\cite{bylander_noise_2011} whose decoherence times have improved from a few nanoseconds to milliseconds in 15 years~\cite{kjaergaard_superconducting_2020}.

In contrast, the characterization of more complex systems such as multilevel nodes and multi-qubit architectures is still a challenge. In particular, the need to characterize correlations between noises affecting different transitions~\cite{sung_multi-level_2021} or different nodes of the quantum architecture~\cite{darrigo_effects_2008} is an emerging issue. A strong motivation of principle is that spatially correlated noise between physical qubits can disrupt one of the pillars of digital quantum computation, namely error correction in logical qubits, apart from the fact that correlated errors are indeed observed in noisy intermediate-scale quantum structures~\cite{vepsalainen_impact_2020}. The effects of spatially correlated low-frequency noise between two quantum devices have been observed for several decades~\cite{ZorinPRB1996background} and their impact on decoherence has been studied~\cite{darrigo_effects_2008} since the first superconducting two-qubit gates were demonstrated. Recently, low-frequency noise correlations between qubit pairs have been characterized by direct measurement of the noise power spectra~\cite{von_lupke_two-qubit_2020,yoneda_noise-correlation_2023,zou_spatially_2023}. However, the complete characterization of the noise acting on a controllable quantum system is a difficult task that requires rapidly increasing resources with the upscaling of the systems. In this work, we attempt to take a different route, which is to capture coarse-grained information starting from the categorization into classes of the multivariable noise acting on a quantum architecture. Our main focus is to find an ML-based method that recognizes the presence of noise correlations. We will see that this possibility is intimately related to the (non-)Markovian character of the dynamics. Correlations be further characterized in a second stage using, among others, dynamical decoupling techniques~\cite{DArrigo2024openloop} useful in the investigation of the effect that non-Markovianity has in the emergence of many-body phenomena~\cite{ChiriacoPRB2023diagrammatic,Tsitsishvili2024measurement}.

For simplicity, we illustrate our ML method for identifying correlations in the paradigmatic case of a three-level network affected by diagonal classical noise. Physically, the model describes a three-site quantum network of single-level quantum dots with tunable tunneling amplitudes~\cite{PopeJSM2019coherent}. The same model describes a qutrit driven by two alternating classical electromagnetic fields. In the former case, we use a control protocol leading to coherent tunneling by adiabatic passage (CTAP)~\cite{GreentreePRB2004coherent,GullansPRB2020coherent}, commonly used for population transfer, as a tool to classify noise. We use supervised learning to identify five classes of noise:
\begin{enumerate}[label=(\arabic*)]
    \item Non-Markovian correlated noise;
    \item Non-Markovian anti-correlated noise;
    \item Non-Markovian uncorrelated noise;
    \item[(4a)] Markovian correlated noise;
    \item[(4b)] Markovian anti-correlated noise.
\end{enumerate}
We show that by measuring the efficiency of CTAP under three different combinations of pulse amplitudes, we can train a neural network to classify noise belonging to four of the five classes mentioned above. In particular, our model can accurately classify noise belonging to the first three classes and Markovian noise (classes 4a and 4b together), while it cannot distinguish between correlated and anti-correlated Markovian noise. Referring to real experiments, we analyze in detail how this information can be extracted from a finite number of projective measurements.

The paper is structured as follows: in Sec.~\ref{sec:noise} we define the models of system and noise that we deal with in our study. In Sec.~\ref{sec:dynamics}
we present the CTAP protocol and the figures of merit that we use in the training phase. In Sec.~\ref{sec:classification} we give a brief overview of \textit{supervised learning} and \textit{neural networks} (NN)~\cite{Burkov2019hundredpage,MarquardtSPLN2021machine,Geron2023handson,Goodfellow2016deep} and describe how we classify among the noises introduced in Sec.~\ref{sec:noise}. Section ~\ref{sec:results} presents the results of the classification for a finite number of samples and imperfect measurements. The possible physical implementations are discussed in Sec.~\ref{sec:physicalimplementation_interpresult} together with a physical interpretation of the results. The conclusions are drawn in Sec.~\ref{sec:conclusion}.

\section{\label{sec:noise}Noisy three-level system}
To fix the ideas we consider a system of three single-level quantum nodes with eigenbasis $\{\ket{i},\; i=0,1,2 \}$ and on-site energies $\mathrm{\epsilon}_i$. This model may describe a system of three quantum dots with an electron
tunneling between them~\cite{GreentreePRB2004coherent}. The tunneling rates between the first and the second dot, $\Op(t)$, and between the second and the third dot, $\Os(t)$, can be controlled by time-dependent external gates. The same model describes a three-level atom driven by two external fields in a multiple rotating frame~\cite{BergmannRMP1998coherent}.

The Hamiltonian of the system is $\Hsys = H_0 + H_\mathrm{c}$, where $H_0$ and $H_\mathrm{c}(t)$ are the free energy and the control term, respectively (throughout the manuscript we use units such that $\hbar=1$), reading
\begin{subequations}\label{eq:Hamiltonian}
    \begin{align}
         & H_0 = \delp\ketbra{1}{1}+ \delta\ketbra{2}{2},                                                                                 \\
         & \Hc = \frac{\Op(t)}{2}\left(\ketbra{0}{1} + \ketbra{1}{0}\right) + \frac{\Os(t)}{2}\left(\ketbra{1}{2} + \ketbra{2}{1}\right),
    \end{align}
\end{subequations}
where, in the language of the three-dot network, we defined the {\it detunings} $\delp = \epsilon_1-\epsilon_0$ and $\delta = \epsilon_2-\epsilon_0$.

We consider diagonal noise, \ie noise affecting the energy levels of the three dots, which may arise from  charge noise, for instance~\cite{yoneda_noise-correlation_2023}. We model it as the following classical stochastic process added to the diagonal entries of the Hamiltonian
\begin{equation}
    \label{eq:Hnoise}
    \Hnoise = \tilde x_1(t)\ketbra{1}{1} + \tilde x_2(t)\ketbra{2}{2}.
\end{equation}
Here, $\tilde x_i(t)$ ($i=1,2$) is a random variable~\cite{Mandel1995optical} with normalized probability density $p(x_i, t)$. 
The total Hamiltonian $H(t)$ in the basis $\{\ket{0}, \ket{1}, \ket{2}\}$ thus reads
\begin{equation}
    \label{eq:totalH}
    H(t) = \Hsys(t) + \Hnoise(t) = \begin{pmatrix}
        0        & \Op(t)/2              & 0                      \\
        \Op(t)/2 & \delp + \tilde x_1(t) & \Os(t)/2               \\
        0        & \Os(t)/2              & \delta + \tilde x_2(t)
    \end{pmatrix}.
\end{equation}
We now link the five classes of noise listed in Sec.~\ref{sec:Intro} to the features of the stochastic mechanism introduced here. We thus consider 
\begin{itemize}
    \item {\bf Non-Markovian types of noise}: When addressing non-Markovian noise, we will make the assumption of {\it quasistatic processes} where the noise mechanism has a long correlation time and can thus be considered constant over the evolution of the system. The  random variables $\tilde x_i(t)$ are assumed to be picked from Gaussian distributions with zero mean, thus describing the cumulative effects of independent  microscopic sources. Needless to say, while $\tilde x_i(t)$ remains constant throughout a single realization of the protocol, it varies between different realizations. We identify the following three classes 
          \begin{enumerate}[label=(\arabic*)]
              \item Correlated: $x_2 (t)= \eta x_1(t)$ with $\eta>0$;
              \item Anti-correlated: $x_2(t) = \eta x_1(t)$ with $\eta<0$;
              \item Uncorrelated: $x_2(t)$ and $x_1(t)$, are independent of each other.
          \end{enumerate}

    \item {\bf Markovian types of noise}: The associated dynamics will be ruled by zero-mean, delta-correlated 
          stochastic processes $\tilde x_i$ making the dynamics of the system dependent only on its current state rather than its past history. We will thus set 
          \begin{equation}
              \label{eq:markovian_correlations}
              \avg{\tilde x_i(t)}=0,\quad \avg{\tilde x_i(t) \tilde x_i(t')} = \gamma\delta(t-t')
          \end{equation}
          and consider the two classes
          \begin{enumerate}
              \item[(4a)] Correlated: $x_2 (t)= \eta x_1(t)$, with $\eta>0$;
              \item[(4b)] Anti-correlated: $x_2(t) = \eta x_1(t)$, with $\eta<0$.
          \end{enumerate}
\end{itemize}

\section{\label{sec:dynamics}Dynamics}
Our method is based on population transfer by CTAP and the closely related stimulated Raman adiabatic passage (STIRAP) scheme, which has found ample use in atomic physics~\cite{BergmannRMP1998coherent,VitanovRMP2017stimulated}.
We shortly review the CTAP/STIRAP approaches and discuss how we solve the equations and then calculate the classification performances in the presence of the different classes of noise considered.
\subsection{CTAP/STIRAP}
CTAP~\cite{GreentreePRB2004coherent} (and its spin equivalent spin-CTAP~\cite{GullansPRB2020coherent}) is a protocol that ideally achieves population transfer from $\ket{0}$ to $\ket{2}$
by adiabatically following a ``trapped state" (often referred to as a dark state) that contains no contribution from the intermediate state $\ket{1}$, which is thus never populated during the protocol. Under ideal conditions, CTAP/STIRAP yields $\sim 100\%$-efficiency population transfer 
with remarkable robustness against parametric fluctuations and state-selectivity.

For the transfer process to be successful it is essential to work at small
detuning, $\delta \ll \Omega_{p,s}$.  In particular,
at the so-called two-photon resonance condition $\delta=0$, the trapped state is an instantaneous eigenstate of the Hamiltonian $\Hsys(t)$ of the form
\begin{equation}
    \label{eq:darkstate}
    \ket{\mathrm{\phi}_\mathrm{D}(t)} = \cos\theta(t)\ket{0}-\sin\theta(t)\ket{2}
\end{equation}
with $\theta(t) = \tan^{-1}\big[\Op(t)/\Os(t)\big]$.
The protocol is operated in the time interval $[\ti,\tf]$ by applying pulses $\Omega_{p,s}(t)$ in a {\it counterintuitive} manner~\cite{BergmannRMP1998coherent}: $\Os$ is applied {\em before} $\Op$, while ensuring that the two pulses overlap in a fraction of the duration of the sequence. In this case, the trapped state $\ket{\phi_\mathrm{D}(t)}$ at the initial time $\ti$ coincides with $\ket{0}$ and at the final time $\tf$ with the target state $\ket{2}$. If the evolution is adiabatic and the system is prepared in $\ket{0}=\ket{\phi_\mathrm{D}(\ti)}$, the system evolves following the trapped state $\ket{\phi_\mathrm{D}(t)}$ throughout the evolution as prescribed by the adiabatic theorem~\cite{BornZFP1928beweis,Messiah1961quantum}. 
The requested adiabaticity of the process can be cast in the form of the global adiabaticity condition~\cite{BergmannRMP1998coherent}
\begin{equation}
    \label{eq:globaladiabaticcondition}
    \Opsmax \tau \geq 10,
\end{equation}
where $\tau$ is the characteristic time scale of the pulses overlap and $\Opsmax = \max_t\Ops(t)$.

\begin{figure}[htbp]
    \centering
    \input{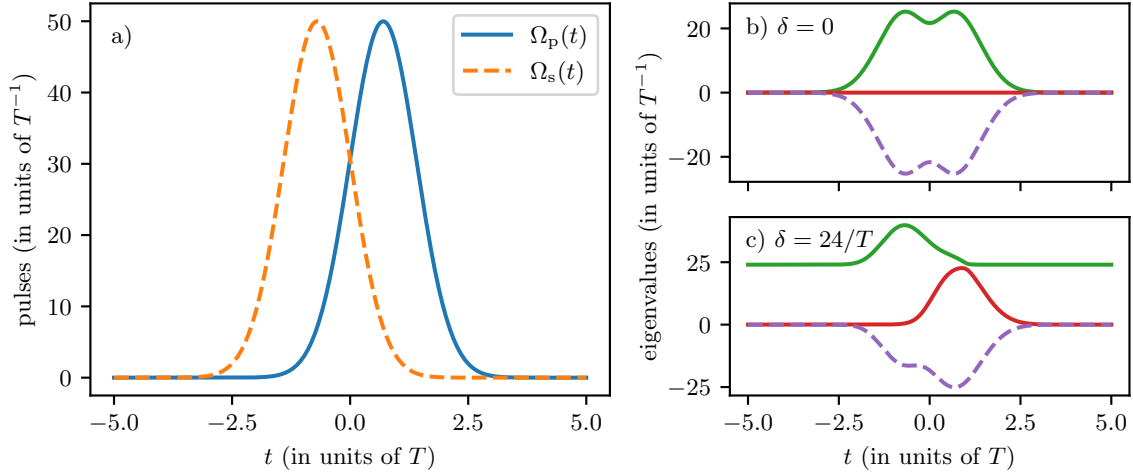}
    \caption{\label{fig:pulses}(a) Example of the pump $\Op(t)$ and Stokes $\Os(t)$ pulses, Eqs.~\eqref{eq:gaussianpulses}, as a function of time $t$. The parameters used are $\Opmax=\Osmax=50/T$, $\tau = 0.7T$ and $t\in[-5T, 5T]$. (b) Instantaneous eigenvalues for $\delta=0$: CTAP/STIRAP achieves 100\% population transfer by adiabatically following the trapped state $\ket{\Phi_D(t)}$ (red line). (c)~Instantaneous eigenvalues for $\delta \neq 0$: The trapped state is not an instantaneous eigenstate and an adiabatic evolution produces only partial population transfer.}
\end{figure}

Under two-photon resonance condition, a counterintuitive sequence satisfying the global adiabaticity request is robust against the actual shape of $\Op$ and $\Os$. In fact, several pulse shapes~\cite{GiannelliPRA2014threelevel} have been used in literature. In this work, we consider Gaussian pulses (see Fig.~\ref{fig:pulses}) of the form
\begin{equation}
    \label{eq:gaussianpulses}
    \Op(t) = \Opmax e^{-(\frac{t-\tau}{T})^2},\qquad \Os(t) = \Osmax e^{-(\frac{t+\tau}{T})^2},
\end{equation}
and we let the system evolve in the time interval $t\in[-5T,5T]$ with $\tau=0.7T$.
We point out that our aim is not achieving an efficient population transfer, but to exploit the sensitivity of the protocol to get information about the noise affecting the system which deteriorates the efficiency. More details about CTAP/STIRAP can be found in~\ref{sec:CTAP}.

\subsection{\label{sec:figuresofmerit}Figures of merit}
The efficiency $\xi$ of the population transfer is defined as the population of the target state $\ket{2}$
at the final time $\tf$, averaged over all the possible realizations of the noise~\cite{Gardiner2010quantum}
\begin{equation}
    \label{eq:efficiency_statavg}
    \xi = \lim_{N\to\infty}\frac{1}{N} \sum_{r=1}^N \abs{\braket{2}{\psii(\tf)}}^2 = \lim_{N\to\infty}\frac{1}{N} \sum_{r=1}^N \xii
\end{equation}
with $\xii = \abs{\braket{2}{\psii(\tf)}}^2$  the population of $\ket{2}$ for a single trajectory $\ket{\psii(t)}$ identified by an individual realization of noise, \ie a choice of $\{x_1^{(r)}(t),x_2^{(r)}(t)\}$. The
index $r=1,\dots,N$ identifies the  trajectory being considered.

The quasistatic noise is constant during an individual trajectory
thus the index $r$ can be unambiguously mapped to the two real values $x_1^{(r)}$ and $x_2^{(r)}$ of the random processes $\tilde{x}_1$ and $\tilde{x}_2$. Therefore $\xii = \xi(x_1^{(r)},x_2^{(r)})$. In this case, the efficiency in Eq.~\eqref{eq:efficiency_statavg} is obtained by averaging over the random processes
\begin{equation}
    \label{eq:efficiency_uncorrelated}
    \xi = \int \xi(x_1,x_2) p(x_1,x_2) \de x_1 \de x_2.
\end{equation}
If the noise is (anti-)correlated, then $x_2^{(r)} = \eta x_1^{(r)}$ 
and Eq.~\ref{eq:efficiency_uncorrelated} simplifies to
\begin{equation}
    \label{eq:efficiency_correlated}
    \xi = \int \xi(x_1,\eta x_1) p_1(x_1) \de x_1.
\end{equation}
For Markovian noise, the average is calculated via the density matrix
$\rho(t)$ of the system which solves a master equation in the Lindblad form (cf.~\ref{sec:markovianlindblad} for the derivation)
\begin{equation}
    \label{eq:lindbaleq}
    \dot\rho(t) = -\I\mathrm[H_0 + H_\mathrm{c}(t), \rho(t)] -\frac{\gamma}{2}(O^2\rho(t) + \rho(t)O^2-2O\rho(t)O),
\end{equation}
where $O = \ketbra{1}{1} + \eta\ketbra{2}{2}$, and $\gamma$ is the dephasing rate given by $\avg{\tilde x_1(t)\tilde x_1(t^\prime)} = \gamma\delta(t-t^\prime)$. The efficiency is given by
\begin{equation}
    \label{eq:efficiencydm}
    \xi =  \bra{2}\rho(\tf) \ket{2}.
\end{equation}

\section{\label{sec:classification}Classification with neural networks}

\subsection{\label{sec:NN}Neural Networks}
Supervised learning requires a labeled dataset $\{(\vect{x}_i, \vect{\hat{y}}_i)\}_{i=1,\dots,N}$, where each input feature vector $\vect{x}_i$ is paired with a target output $\vect{\hat{y}}_i$. To perform classification, we employ a feedforward neural network, specifically a multilayer perceptron (MLP).

An MLP consists of an input layer, one or more hidden layers, and an output layer. Each neuron computes its output by applying a nonlinear activation function to a weighted sum of the outputs from the previous layer:
\begin{subequations}
    \begin{equation}
        \label{eq:yl}
        \vect{y}^{(l)} = f^{(l)}(\vect{z}^{(l)}),
    \end{equation}
    where $f^{(l)}$ is a nonlinear funtion called the \textit{activation function}, and
    \begin{equation}
        \label{eq:zl}
        \vect{z}^{(l)} = \vect{w}^{(l)}\vect{y}^{(l-1)} + \vect{b}^{(l)}.
    \end{equation}
\end{subequations}
Here $\vect{w}^{(l)} \in \mathbb{R}^{D^{(l)} \times D^{(l-1)}}$ is the weight matrix, $\vect{b}^{(l)} \in \mathbb{R}^{D^{(l)}}$ is the bias vector, and $f^{(l)}$ is the activation function for layer $l$. The input layer values are set to $\vect{y}^{(0)} = \vect{x}$, and the final output is $\vect{y} = \vect{y}^{(L)}$.

Training the network involves finding weights $\vect{W} = \{\vect{w}^{(l)}\}_l$ and biases $\vect{B} = \{\vect{b}^{(l)}\}_l$ that minimize a cost function $C(\{\vect{y}_i, \vect{\hat{y}}_i\}_i | \vect{W}, \vect{B})$, which quantifies the error between the predicted outputs $\vect{y}_i$ and the target outputs $\vect{\hat{y}}_i$. Typically, this minimization process is executed using algorithms such as \textit{stochastic gradient descent} or its variants. The  gradients needed for these optimization algorithms are computed through backpropagation, a highly efficient algorithm that leverages the chain rule to propagate errors backward through the network.

We use a neural network consisting of two hidden layers with 128 and 100 neurons, respectively, as detailed in Table~\ref{tab:NNstructure}. For the hidden layers, we employ the \textit{leaky rectified linear unit} (LeakyReLU)~\cite{Maas2013rectifier} as the activation function, defined as:
\begin{equation}
    \label{eq:LeakyRELU}
    f_{\text{LReLU}}(z) =
    \begin{cases}
        z        & \text{if} \quad z \geq 0, \\
        \alpha z & \text{if} \quad z < 0,
    \end{cases}
\end{equation}
where $\alpha$ is set to $0.01$. This activation function helps to mitigate the ``dying ReLU'' problem~\cite{Douglas25ACSSC2018why}, where neurons become inactive and only output zero.
\begin{table}[!b]
    \renewcommand{\arraystretch}{1.4}
    \centering
    \begin{tabular}{c|c|c}
        \toprule
        \tableheadline{Layer} & \tableheadline{$\#$ neurons} & \tableheadline{Activation function} \\ \midrule
        input                 & $3$                          &                                     \\ \hline
        hidden 1              & $128$                        & LeakyReLU                           \\ \hline
        hidden 2              & $100$                        & LeakyReLU                           \\ \hline
        output                & $4$ or $5$                   & Softmax                             \\
        \bottomrule
    \end{tabular}
    \caption{\label{tab:NNstructure}Structure of the neural network used for classification of the 4 or 5 classes of noise.}
\end{table}

For the output layer, we use the \textit{softmax} activation function to obtain probabilities over the classes:
\begin{equation}
    \label{eq:softmax}
    \vect{y} = f_\textrm{softmax}(\vect{z}^{(L)}) = \frac{e^{\vect{z}^{(L)}}}{\sum_{k=1}^{D^{(L)}} e^{z_k^{(L)}}},
\end{equation}
where the exponentiation and division are performed element-wise. This ensures that the output vector $\vect{y}$ sums to one, allowing interpretation as class probabilities. We use one-hot encoding for the target outputs $\vect{\hat{y}}_i$, where each class is represented by a binary vector with a single one corresponding to the correct class.

The network is trained by minimizing the \textit{categorical cross-entropy} loss function:
\begin{equation}
    \label{eq:categorical_cross-entropy}
    C(\{\vect{y}_i, \vect{\hat{y}}_i\}_i) = -\frac{1}{N}\sum_{i=1}^N \sum_{j=1}^{D^{(L)}} \hat{y}_{ij} \ln(y_{ij}),
\end{equation}
where $y_{ij}$ is the predicted probability for class $j$ of sample $i$, and $\hat{y}_{ij}$ is the corresponding element of the one-hot encoded target vector.

\subsection{\label{sec:data_generation}Data generation}
\begin{figure}[!htbp]
    \centering
    \includegraphics[width=0.9\textwidth]{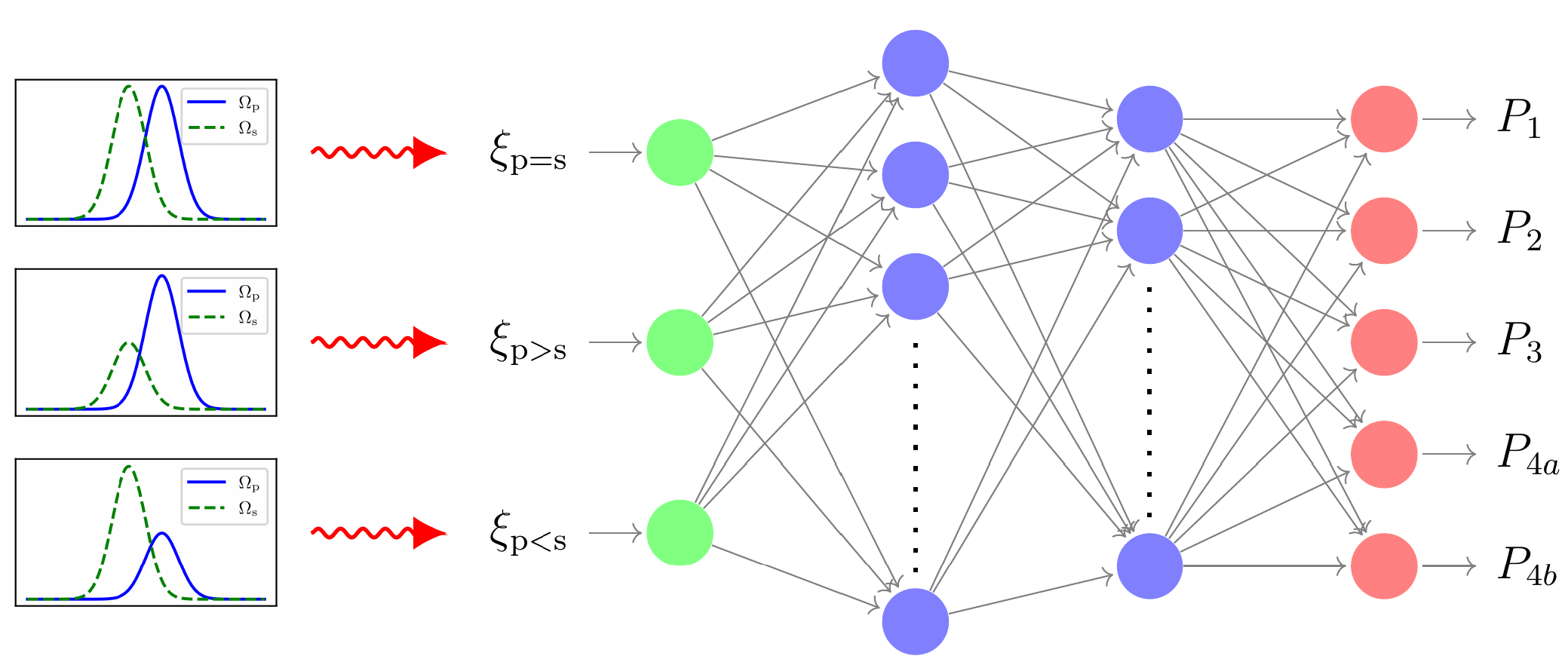}
    \caption{\label{fig:NNscheme}Schematic representation of our model. The efficiencies obtained with the three driving conditions $\Opmax = \Osmax$, $\Opmax > \Osmax$, and $\Opmax < \Osmax$ are given as input to the neural network. The Neural Network is composed of 2 hidden layers with $128$ and $100$ neurons respectively, and activation function LeakyReLU, Eq.~\eqref{eq:LeakyRELU}. The output layer composed of $4$ or $5$ neurons represent the probability of assigning each data to the relative noise class. This is achieved with the softmax activation function, Eq.~\eqref{eq:softmax}, on the output layer.}
\end{figure}

In order to efficiently classify between the different noise types it is imperative to identify features which are sensitive to the distinct characteristics of the classes introduced in Sec.~\ref{sec:noise}.
To this end, we choose as features the efficiency of the CTAP/STIRAP protocol under three different driving conditions. Notice that in this approach, in contrast to other works where the time series of expectation values of several operators are used as input~\cite{BarrMLST2024spectral,GuoAPL2023highspeed,ZengAPL2021application}, we only need the expectation value of the operator $\ketbra{2}{2}$ at the final time of the evolution. This makes our method more efficient and appealing for experimental applications.
Referring to Eqs.~\eqref{eq:gaussianpulses} we consider:
\begin{enumerate*}[]
    \item $\Opmax = \Osmax$,
    \item $\Opmax > \Osmax$,
    \item $\Opmax < \Osmax$.
\end{enumerate*}
For each type of noise,
we solve numerically the dynamics of the system under the three driving conditions and compute the efficiency as described in Sec.~\ref{sec:dynamics}.
We denote the efficiencies as $\xieq$, $\xigr$, and $\xilo$, for the driving conditions (i), (ii), and (iii) respectively,  and we use $\vect{x} = (\xieq, \xigr, \xilo)$ as input feature vector to the NN, see Fig.~\ref{fig:NNscheme}.
The parameters we use are (i) $\Opmax = \Osmax = 50/T$, (ii) $\Opmax=20\sqrt{10}/T > \Osmax=10\sqrt{10}/T$, and (iii) $\Opmax=10\sqrt{10}/T < \Osmax=20\sqrt{10}/T$ with evolution time $t\in[-5T,5T]$, $\tau=0.7T$ and $\sigma_1\approx 17.6/T$ being the standard deviation of the Gaussian distribution $p_1(x_1,t)$.

For the correlated and anti-correlated noise, data are generated by randomly sampling the correlation parameter $\eta$ in the intervals $[0.1,5]$ and $[-5,-0.1]$, respectively. For each randomly selected $\eta$ we calculate three efficiencies, one for each of the pulse conditions (i), (ii), and (iii), thus generating the input feature vector $\vect{x}$. The efficiencies are calculated using Eq.~\eqref{eq:efficiency_correlated} for non-Markovian quasistatic noise (with $p_1(x_1,t)$ being a Gaussian distribution) and Eq.~\eqref{eq:efficiencydm} for Markovian noise.

For uncorrelated, non-Markovian quasistatic noise, the values of $\tilde x_1$ and $\tilde x_2$ are independently sampled from two Gaussian probability distributions $p_1(x_1,t)$ and $p_2(x_2,t)$ with mean $\mu_1=\mu_2=0$. The standard deviation $\sigma_1$ of the first distribution is kept fixed, while the standard deviation $\sigma_2$ of the distribution $p_2(x_2,t)$ is randomly sampled in the interval $[-5\sigma_1, 5\sigma_1]$. As mentioned in Sec.~\ref{sec:NN}, the classes are one-hot encoded, thus the output of the neural network is a layer with four or five neurons. For each noise class, we generate $500$ samples such that the total data consists of $2000$ or $2500$ data points $\{\vect{x}, \vect{\hat{y}}\}$. We then split the data in a training, validation and test set with a ratio of $0.6:0.2:0.2$, respectively.

In order to evaluate the efficacy of the model we use the accuracy $A$ defined as the number of correct predictions divided by the number of total predictions $N$
\begin{equation}
    \label{eq:accuracy}
    A = \frac{1}{N}\sum_{i=1}^N \delta (\argmax{\vect{y}_i}, \argmax{\vect{\hat y}_i})
    ,
\end{equation}
where the $\argmax$ function yields the index $j$ of the maximal component of $\vect{y}$ and $\delta(\cdot)$ is the Kronecker delta.

\section{\label{sec:results}Results}
\begin{figure}[htbp]
    \centering
    \input{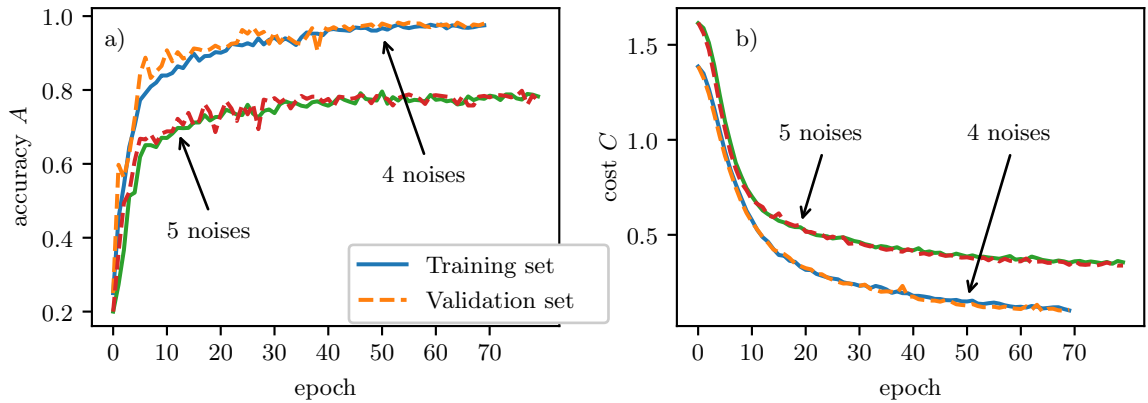}
    \caption{\label{fig:ML_results_4_and_5noises}(a) Accuracy, Eq.~\eqref{eq:accuracy}, and (b) value of the cost function, Eq.~\eqref{eq:categorical_cross-entropy}, for the training (solid blue for four noises and solid green for five noises) and validation (dashed orange for four noises and dashed red for five noises) sets versus the number of epochs of training. The accuracy on the test set is $A\approx1$ for four noise classes and $A\approx0.81$ for five noise classes.}
\end{figure}

\begin{figure}[htbp]
    \centering
    \includegraphics{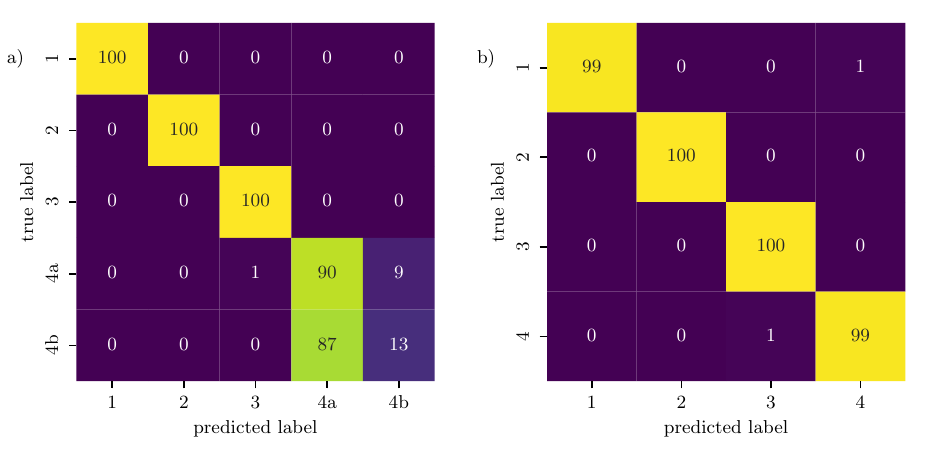}
    \caption{Confusion matrix for the MLP model for the classification of the classes of noise considered in this work. Each row of the matrix represents the instances of the true classes, while each column represents the instances of the predicted classes. It is apparent that classes $4a$ and $4b$, \ie Markovian correlated and anti-correlated, are not easily distinguishable.}
    \label{fig:confusion_matrix_5noises}
\end{figure}

The ML model and the training are performed with TensorFlow~\cite{tensorflow2015-whitepaper} and the results are shown in Fig.~\ref{fig:ML_results_4_and_5noises}. The accuracy $A$ for the training and validation sets is shown in Fig.~\ref{fig:ML_results_4_and_5noises}(a). After training on the test set the accuracy of the model is $A\approx0.81$ and varies within the values of approx $0.79$ and $0.81$ depending on the random initialization of the NN weights and the random shuffling of the data for the splitting in the three sets. For the same test set, we report in Fig.~\ref{fig:ML_results_4_and_5noises}(b) how the value of the cost function $C$ changes during training.
In Fig.~\ref{fig:confusion_matrix_5noises}, we show the {confusion matrix} for the trained model, a tabular representation used to evaluate the performance of a classification model. It cross-tabulates the actual class labels with the model's predictions, providing insight into the correct and incorrect classifications made by the model. Thus it helps to identify which classes are not easily distinguishable from each other posing a challenge to the classification. Each row of the matrix represents the instances of the actual classes, while each column represents the instances of the predicted classes. It is apparent from Fig.~\ref{fig:confusion_matrix_5noises}(a) that classes $4a$ and $4b$ are not distinguishable from each other with the input features of our choice and almost all the samples collapse in class $4a$.

We then repeat the same analysis considering the Markovian noise as a single class, \ie by grouping together correlated and anti-correlated noise and taking $\eta\in[-5,5]$. As expected, the effectiveness of the model increases reaching an accuracy $A\approx1$ that varies between $\simeq0.97$ and $1$ depending on the random initialization. The accuracy $A$ and the value of the cost function $C$ during the training are shown in Fig.~\ref{fig:ML_results_4_and_5noises}. The confusion matrix for this model is reported in Fig.~\ref{fig:confusion_matrix_5noises}(b) showing that now the four classes are clearly distinguishable with the chosen input features.

\subsection{\label{sec:results_finitenmeasurement}Results for a finite number of measurement}
The analysis performed in the previous section describes the ideal physical situation that the data are obtained from an infinite number of $100\%$-efficiency of projective quantum measurements. While very large efficiency can be achieved by quantum non-demolition measurements~\cite{braginsky_quantum_1992}, in this section we analyze how
the training and the accuracy of the model are affected by the number of measurements $M$ being finite.
\begin{figure}[!b]
    \centering
    \input{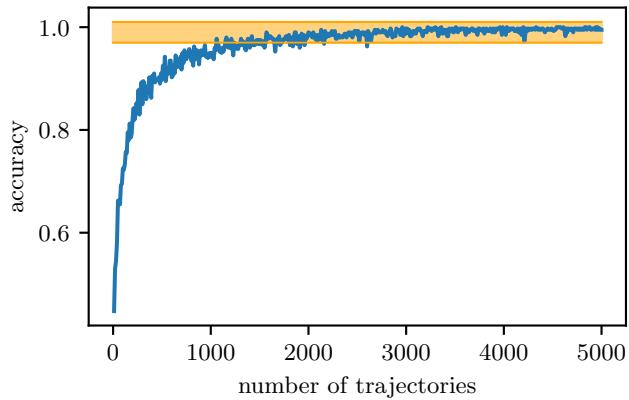}
    \caption{Accuracy of the classification task for four classes of noise versus the number $M$ of projective measurements on quantum trajectories. The shaded orange region corresponds to the accuracy obtained with ideal measurements.}
    \label{fig:accuracy_vs_trajectories}
\end{figure}
To this end, we employ the same method described in Sec.~\ref{sec:data_generation} but compute the efficiency $\xi_M$ as the one resulting from
a set of individual ``quantum" trajectories. The result
of unravelling each trajectory is simulated
by extracting each time the value $1$ with probability equal to the population of the target state $\ket{2}$.  The output $\vect{y}$ is produced by
averaging over the $M$ quantum trajectories. We produce $500$ distinct datasets, each corresponding to a different measurement count $M=10,\dots,5000$ (in steps of $10$). For each $M$, we train the model and evaluate its accuracy using the same methodology as previously described.

The results are presented in Fig.~\ref{fig:accuracy_vs_trajectories}, where we report the accuracy versus the number of measurements. As expected, the accuracy improves for increasing $M$, approaching the value obtained in the ideal case,
which lies in the range $A\approx[0.97,1]$.

\section{\label{sec:physicalimplementation_interpresult}Physical implementation and interpretation of the results}
\subsection{Physical implementations of the model}
Spatially correlated noise in solid-state quantum information processing is a topic of current experimental investigation in superconducting~\cite{von_lupke_two-qubit_2020} and semiconducting~\cite{zou_spatially_2023} quantum devices. Charge- and spin-CTAP have been studied extensively~\cite{menchon-enrich_spatial_2016,GullansPRB2020coherent} and recently a similar protocol has been implemented in semiconducting quantum dots~\cite{kandel_adiabatic_2021} while CTAP of electromagnetic excitations has been proposed as the building block of quantum operations in the ultrastrong coupling regime~\cite{falci_ultrastrong_2019,giannelli_detecting_2024} circuit-QED architectures.  In superconducting devices, STIRAP has been demonstrated in Vee configuration~\cite{kumar_stimulated_2016, xu_coherent_2016,GongPRA2024twophotontransition} whereas the Lambda configuration we study in this work could be implemented either directly~\cite{siewert_advanced_2009,falci_design_2013,earnest_realization_2018} or by a detuning-modulated protocol~\cite{di_stefano_coherent_2016} which generalizes hyper-STIRAP~\cite{BergmannRMP1998coherent} and bypasses parity selection rules at noise-protected operating points.

The model of noise studied in this paper accounts for the main effect of sources inducing a fluctuating electric or magnetic polarization of the device. In particular, noise coupled by an operator which commutes with the uncoupled (for CTAP) or undriven (for STIRAP) Hamiltonian is called ``longitudinal'' and determines stochastic fluctuations of the bare energy splittings of the devices. In principle, the noise also produces ``transverse'' fluctuations of the off-diagonal entries of the Hamiltonian which affect the splittings only at second-order. Moreover, CTAP and STIRAP have the remarkable property that the efficiency of population transfer is almost insensitive to transverse parametric fluctuations~\cite{VitanovRMP2017stimulated} which therefore can be ignored. As for the correlations, in CTAP they originate from microscopic noise sources acting on the network in a spatially correlated manner. For STIRAP in a multilevel artificial atom, since noise couples to the device through a single operator it induces fluctuations of the energy level related to each other. In this latter case, the $\eta$ is given by the ratio of the first derivatives of the energy spectrum at the given bias.

\subsection{\label{sec:resultsinterpretation}Interpretation of the results}
\begin{figure*}[htbp]
    \centering
    \input{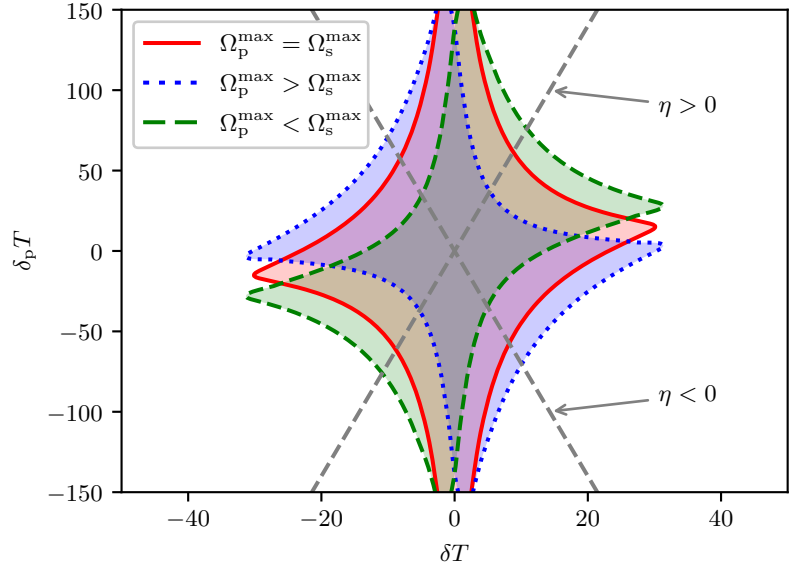}
    \caption{\label{fig:stability_plots}Efficiency of CTAP for the three driving conditions (i) $\Opmax = \Osmax = 50/T$ (red, solid line), (ii) $\Opmax=20\sqrt{10}/T > \Osmax=10\sqrt{10}/T$ (blue, dotted line), and (iii) $\Opmax=10\sqrt{10}/T < \Osmax=20\sqrt{10}/T$ (green dashed line). The shaded areas within the boundary lines highlight the regions, in the parameter space, where the efficiency is $>0.7$ (the efficiency being exactly 0.7 along the boundary curves). 
        The efficiencies are calculated with Eqs.~\eqref{eq:efficiency_uncorrelated} and~\eqref{eq:efficiency_correlated}. The dashed gray lines represent correlation ($\eta>0$) or anti-correlation ($\eta<0$) between the detunings $\delta$ and $\delp$. The other parameter is $\tau=0.7T$.}
\end{figure*}

We now seek an explanation of the ability to classify among differently correlated non-Markovian quasistatic noises. To this end, we analyze the stability plots (efficiency versus the detunings $\delta$ and $\delp$) for the three driving conditions, shown in Fig.~\ref{fig:stability_plots}

Ideally, the system is operated at zero detunings, $\delta=\delta_p=0$.
The effect of quasistatic noise is to move, at each individual repetition of the protocol, the point associated with the operations of the system to somewhere else in the space of parameters [cf. Fig.\ref{fig:stability_plots}]. The efficiency for correlated and anti-correlated noise types, is calculated as a weighed average over the line $\delta = \eta \delp$, see Eq.~\eqref{eq:efficiency_correlated}. The  corresponding efficiency depends on the ratio between $\Opmax$ and $\Osmax$, this dependence being different for $\eta>0$ or $\eta<0$. Fig.~\ref{fig:stability_plots} reports as an example the lines $\delta=\eta\delp$ for $\eta\gtrless0$: it is clear that for $\eta>0$ the efficiency of CTAP with $\Opmax<\Osmax$ is higher than the one with $\Opmax>\Osmax$, and vice versa for $\eta<0$. For a more extensive analysis of this behavior we refer to Ref.~\cite{FalciPS2012effects}. For uncorrelated noise, instead, the average is over the whole $\delta-\delp$ plane -- cf. Eq.~\eqref{eq:efficiency_uncorrelated} -- and its dependence on the three different driving conditions is not as explicit.

Such analysis cannot be replicated for Markovian noise, as individual trajectories cannot be represented as single points in Fig.~\ref{fig:stability_plots}. Instead, the noise assumes all values along the line $\delta=\eta\delp$ during each repetition of the protocol.

\section{\label{sec:conclusion}Conclusions}
In this work, a ML-based method for categorizing correlated classical noise affecting a three-level quantum system has been presented. We have considered different cases of non-Markovian and Markovian noise acting on a multilevel system focusing on recognizing correlations in the induced fluctuations of the energy splittings.

We have shown that the sensitivity of a CTAP/STIRAP protocol to correlation and Markovianity allow to classify the noise main characteristics in great detail. In particular, data on the efficiency of population transfer under three different driving conditions is sufficient to distinguish between the correlations of non-Markovian quasi-static noise and to discriminate between Markovian and non-Markovian noise. In contrast, this approach does not recognize correlations of Markovian noises. As input to the ML analysis, we have used the numerical solution of the stochastic Schr\"odinger equation and of the Markovian Lindblad-Master equation, which provide the averages over an infinite number of measurements. Our study was complemented by simulating real projective quantum measurements showing that our approach is robust against errors due to the finite statistics. We stress that the implications of our work go beyond the classification of noise. The methodology developed here provides a new perspective on the use of quantum protocols such as CTAP/STIRAP as a tool for environmental diagnostics.

Future directions of this research include exploring the potential of incorporating alternative features of the input to extend the approach to discriminate other classes of noise and to unsupervised learning. In addition, we will investigate the interplay with relaxation~\cite{falci_design_2013, spagnolo_relaxation_2012} and extend the diagnostics to more complex probes, such as interacting multiqubit systems~\cite{yoneda_noise-correlation_2023}, quantum dot chains~\cite{mills_shuttling_2019}, strongly~\cite{blais_quantum_2020} and ultrastrongly coupled~\cite{falci_ultrastrong_2019,giannelli_detecting_2024} circuit-QED architectures. In these cases control could be implemented by unconventional CTAP/STIRAP protocols operated by detunings~\cite{di_stefano_population_2015,BrownNJP2021reinforcement}. This work and the foreseen advances could provide important elements for developing new integrated robust quantum control and error correction strategies.

\section*{Data availability statement}
The code and the data that support the findings of this study are openly available at the following URL/DOI: \url{https://github.com/Shreyasi31/SmallQNetNoiseML}.

\section*{Acknowledgments}
The authors acknowledge Dario Allegra, Lorenzo Catania, Vincenzo Minissale and Giuliano Chiriac\`o for useful discussions.  SM acknowledges support from the project PON Ricerca e Innovazione 2014-2020, React EU, Axis IV, Action IV.4 (CUP: E69J21011430006), within the PhD course in Physics of the Catania University. LG and EP acknowledge support from the PNRR MUR project PE0000023-NQSTI; GF is supported by the  ICSC – Centro Nazionale di Ricerca in High-Performance Computing, Big Data and Quantum Computing. GF and EP acknowledge support from the University of Catania, Piano Incentivi Ricerca di Ateneo 2020-22, project Q-ICT. EP acknowledges the COST Action  SUPERQUMAP (CA 21144). MP acknowledges support from the European Union's Horizon Europe EIC-Pathfinder project QuCoM (grant no. 101046973), the Royal Society Wolfson Fellowship (grant no. RSWF/R3/183013), the UK EPSRC (grant no. EP/T028424/1), and the Department for the Economy Northern Ireland under the US-Ireland R\&D Partnership Programme.

\clearpage

\appendix

\section{\label{sec:markovianlindblad}Derivation of the Lindblad master equation for Markovian diagonal classical noise}
The Lindblad master equation Eq.~\eqref{eq:lindbaleq} describing of the dynamics of the system subject to Markovian noise is derived as follows. We start from the total Hamiltonian of the system $H = \Hsys + \Hnoise$, where $\Hnoise$ is given by Eq.~\eqref{eq:Hnoise}. Using the correlation $x_2(t) = \eta x_1(t)$, we write
\begin{equation}
    \Hnoise = \tilde x_1(t)\left(\ketbra{1}{1} + \eta\ketbra{2}{2}\right) = \tilde x_1(t)O,
\end{equation}
where $O = \ketbra{1}{1} + \eta\ketbra{2}{2}$ is the noise operator. We assume that noise has zero mean $\avg{\tilde x_1(t)} = 0$  and it is Markovian, i.e. $\avg{\tilde x_1(t) \tilde x_1(t^\prime)} = \hbar^2\gamma\delta(t-t^\prime)$. Since the dynamics of the system is adiabatic and the noise has vanishing correlation time, we can write a master equation for the time-independent system dynamics for each $t$. This means that we can derive the master equation in an ``instantaneous interaction picture''.
\begin{equation}
    \tilde{\rho}(t) = e^{\frac{\I}{\hbar}\Hsys t}\rho(t)e^{-\frac{\I}{\hbar}\Hsys t},
\end{equation}
where $\rho(t)$ is the density operator in the Schr\"odinger picture. Physically this approximation amounts in neglecting the backaction of the environment due to the non-adiabatic transitions of the system.
The time evolution of the system is thus given by
\begin{equation}
    \label{eq:Vonneumann_intpic}
    \dot{\tilde{\rho}}(t)= -\frac{\I}{\hbar}[\Hnoisetilde (t), \tilde{\rho}(t)],
\end{equation}
with $\Hnoisetilde(t) = e^{\frac{\I}{\hbar}\Hsys t}\Hnoise (t)e^{-\frac{\I}{\hbar}\Hsys t}$.
Integrating over the time interval $[0, t]$, we get
\begin{equation}
    \tilde{\rho}(t)= \tilde{\rho}(0)-\frac{\I}{\hbar}\int_0^{t}[\Hnoisetilde (t'), \tilde{\rho}(t')]dt'.
\end{equation}
Substituting $\tilde{\rho}(t)$ in Eq.~\eqref{eq:Vonneumann_intpic} we obtain
\begin{equation}
    \dot{\tilde{\rho}}(t)= -\frac{\I}{\hbar}e^{\frac{\I}{\hbar}\Hsys t}[\tilde x_1(t)O\rho(0) - \rho(0)O \tilde x_1(t)]e^{-\frac{\I}{\hbar}\Hsys t} - \frac{1}{\hbar^2}\int_0^t[\Hnoisetilde (t), [\Hnoise (t^\prime), \tilde{\rho}(t^\prime)]]dt^\prime.
\end{equation}
Averaging over the stochastic variable at time $t$, we obtain
\begin{equation}
    \langle \dot{\tilde{\rho}}(t) \rangle =  - \frac{1}{\hbar^2}\biggl< \int_0^t[\tilde x(t)\tilde O(t), [\tilde x(t^\prime)\tilde O(t^\prime), \tilde{\rho}(t^\prime)]]dt^\prime \biggr>,
\end{equation}
where, $\tilde O (t) = e^{\frac{\I}{\hbar}\Hsys t}Oe^{-\frac{\I}{\hbar}\Hsys t}$. Taking into account that $t^\prime < t$ and $\tilde \rho(t^\prime)$ cannot depend on $t$ except for $t^\prime =t$, we can write
$\avg{\tilde x(t)\tilde x(t^\prime)\tilde\rho(t^\prime)}=\avg{\tilde x(t)\tilde x(t^\prime)}\tilde\rho(t^\prime)$. We thus obtain
\begin{equation}
    \langle \dot{\tilde{\rho}}(t) \rangle = -\frac{\gamma}{2}\left(\tilde O^2(t)\tilde{\rho}(t) + \tilde{\rho}(t)\tilde O^2(t) - 2 \tilde O(t)\tilde{\rho}(t)\tilde O(t)\right).
\end{equation}
As expected, this equation can be recast in the Lindblad form. In the Schr\"odinger picture we obtain
\begin{equation}
    \langle \dot{\rho}(t) \rangle = -\frac{\I}{\hbar}[\Hsys, \rho(t)] - \frac{\gamma}{2}\left(O^2\rho(t) + \rho(t)O^2-2O\rho(t)O\right).
\end{equation}

\section{\label{sec:CTAP}Adiabatic passage in a three-node network}
We consider the Hamiltonian of the three-node network $\Hsys(t) = H_0 + H_\mathrm{c}(t)$, where $H_0$ and $H_\mathrm{c}(t)$ are the uncoupled and the control parts, respectively ($\hbar=1$):
\begin{subequations}\label{eq:Hamiltonian_app}
    \begin{align}
         & H_0 = \delp\ketbra{1}{1}+ \delta\ketbra{2}{2},                                                                                 \\
         & \Hc = \frac{\Op(t)}{2}\left(\ketbra{0}{1} + \ketbra{1}{0}\right) + \frac{\Os(t)}{2}\left(\ketbra{1}{2} + \ketbra{2}{1}\right),
    \end{align}
\end{subequations}
and where $\delp = \epsilon_1-\epsilon_0$ and $\delta = \epsilon_2-\epsilon_0$ define the detunings. CTAP~\cite{GreentreePRB2004coherent} is a protocol that achieves population transfer from $\ket{0}$ to $\ket{2}$  by adiabatic following a trapped state and never populating the intermediate state $\ket{1}$. The atomic analogue is  STIRAP~\cite{BergmannRMP1998coherent,VitanovRMP2017stimulated} where high-fidelity population transfer occurs via a dark state. These protocols have remarkable robustness against parametric fluctuations.

CTAP is based on the counter-intuitive ordering of the pulses $\Op(t)$ and $\Os(t)$. The condition $\delta\approx0$ is important for the successful population transfer~\cite{VitanovRMP2017stimulated}. For $\delta=0$, the instantaneous eigenstates of $\Hsys(t)$ are
\begin{subequations}
    \label{eq:dresses_states}
    \begin{align}
         & \ket{\mathrm{\phi}_\mathrm{D}(t)} = \cos\theta(t)\ket{0}-\sin\theta(t)\ket{2},                                  \\
         & \ket{\mathrm{\phi}_-(t)} = \sin\theta(t)\cos\phi(t)\ket{0}-\sin\phi(t)\ket{1}+\cos\theta(t)\cos\phi(t)\ket{2},  \\
         & \ket{\mathrm{\phi}_+(t)} = \sin\theta(t)\sin\phi(t)\ket{0}-\cos\phi(t)\ket{1} +\cos\theta(t)\sin\phi(t)\ket{2},
    \end{align}
\end{subequations}
with eigenvalues
\begin{subequations}
    \label{eq:instanteigenvalues}
    \begin{align}
         & \lambda_\mathrm{D}(t) = 0,                                          \\
         & \lambda_-(t) = -\frac{\hbar}{2}\sqrt{\Op(t)^2+\Os(t)^2}\tan\phi(t), \\
         & \lambda_+(t) = \frac{\hbar}{2}\sqrt{\Op(t)^2+\Os(t)^2}\cot\phi(t),
    \end{align}
\end{subequations}
where
\begin{align}
     & \theta(t) = \tan^{-1}\left(\frac{\Op(t)}{\Os(t)}\right),                                                 \\
     & \phi(t) = \tan^{-1}\left(\frac{\sqrt{\Op(t)^2+\Os(t)^2}}{\delp+\sqrt{\delp^2+\Op(t)^2+\Os(t)^2}}\right).
\end{align}

The eigenstate $\ket{\phi_\mathrm{D}(t)}$ corresponding to zero eigenvalue is called the \textit{trapped state} since it is trapped in the subspace $\{\ket{0},\ket{2}\}$. For STIRAP in multilevel atoms it is called \textit{dark state} since it cannot absorb or emit photons despite the external driving.  The states $\ket{\phi_-(t)}$ and $\ket{\phi_+(t)}$ are called Autler-Townes states and $|\lambda_+-\lambda_-|$ the Autler-Townes splitting.

If the pulses are counter-intuitively ordered, i.e. if $\Os$ is applied before $\Op$:
\begin{equation}
    \label{eq:counterintuitiveorder}
    \lim_{t\to\ti} \frac{\Op(t)}{\Os(t)} = \lim_{t\to\tf} \frac{\Os(t)}{\Op(t)} = 0,
\end{equation}
then the dark state $\phi_\mathrm{D}(t)$ at the initial time $\ti$ will coincide with the initial state $\ket{0}$ and at the final time $\tf$ will coincide with the target state $\ket{2}$. If the evolution is adiabatic and the system is initially prepared in $\ket{0}=\ket{\phi_\mathrm{D}(\ti)}$, by the adiabatic theorem~\cite{BornZFP1928beweis,Messiah1961quantum}, the system evolves following the dark state $\ket{\phi_\mathrm{D}(t)}$ throughout the time evolution. Thus, the population is transferred from state $\ket{0}$ to state $\ket{2}$ never populating the intermediate state $\ket{1}$. The adiabaticity condition is given by~\cite{GreentreePRB2004coherent,BergmannRMP1998coherent}
\begin{equation}
    \label{eq:adiabaticcondition}
    \hbar\abs{\braket{\phi_{\pm}}{\mathrm{\dot{\phi}_\mathrm{D}}}}\ll\abs{\lambda_0-\lambda_{\pm}}.
\end{equation}
Using the expressions for the dressed basis, Eqs.~\eqref{eq:dresses_states} and Eqs.~\eqref{eq:instanteigenvalues}, equation~\eqref{eq:adiabaticcondition} is recast in terms of the pulses~\cite{BergmannRMP1998coherent,VitanovRMP2017stimulated,GiannelliPLA2022tutorial}
\begin{equation}
    \label{eq:localadiabaticitycondition}
    \abs{\dot{\theta}(t)} \ll \frac{1}{2}\absbr{\delp\pm\sqrt{\delp^2+\Op(t)^2+\Os(t)^2}},
\end{equation}
yielding the so-called ``local adiabaticity condition'' which has to hold for all times $t\in[\ti,\tf]$. By time averaging Eq.~\eqref{eq:localadiabaticitycondition} over $\tau$ and assuming $\delp\ll\Opmax,\Osmax$ one can obtain the weaker ``global adiabaticity condition'' which is often reported as~\cite{BergmannRMP1998coherent}
\begin{equation}
    \label{eq:globaladiabaticcondition2}
    \Opsmax \tau \geq 10,
\end{equation}
where $\tau$ is the characteristic time scale of the pulses overlap and $\Opsmax = \max_t\Ops(t)$.

In this work, we perform the population transfer using Gaussian pulses of the form
\begin{subequations}
    \label{eq:gaussianpulses_app}
    \begin{align}
         & \Op(t) = \Opmax e^{-(\frac{t-\tau}{T})^2}, \\
         & \Os(t) = \Osmax e^{-(\frac{t+\tau}{T})^2},
    \end{align}
\end{subequations}
and we let the system evolve in the time interval $t\in[-5T,5T]$ with $\tau=0.7T$, see Fig.~\ref{fig:pulses} for an example of pulses.

\clearpage

\printbibliography

\end{document}